\documentclass[conference]{IEEEtran}
\IEEEoverridecommandlockouts

\usepackage{xcolor}
\usepackage[pdftex]{graphicx} % pdftex dvips dvipsone dvipdf
	\graphicspath{{./}}
\usepackage{caption}
\usepackage{subcaption}
\usepackage{wrapfig}

\usepackage[cmex10]{amsmath}
\usepackage{amsfonts}

\usepackage{array}
\usepackage{multirow}

\DeclareMathOperator*{\argmin}{arg\,min}

\def\BibTeX{{\rm B\kern-.05em{\sc i\kern-.025em b}\kern-.08em
    T\kern-.1667em\lower.7ex\hbox{E}\kern-.125emX}}

\newcommand{\eucnorm}[1]{\ensuremath{\|#1\|}}

\newcommand{\hex}[1]{\ensuremath{\texttt{0x#1}}}

\renewcommand{\vec}{\boldsymbol}

\begin{document}

\thispagestyle{empty}

\begin{figure*}[!t]\large
This paper is a preprint; it has been accepted for publication in 2019 IEEE International Conference on Communications (IEEE ICC), 20--24 May 2019, Shanghai, China.
\medskip

{\bf IEEE copyright notice}
\smallskip

\copyright\ 2019 IEEE. Personal use of this material is permitted. Permission from IEEE must be obtained for all other uses, in any current or future media, including reprinting/republishing this material for advertising or promotional purposes, creating new collective works, for resale or redistribution to servers or lists, or reuse of any copyrighted component of this work in other works.
\vspace*{300pt}

\mbox{~}
\end{figure*}

\newpage

\title{A Novel Malware Detection System Based On Machine Learning and Binary Visualization%
\thanks{%
\protect\begin{wrapfigure}[3]{l}{.9cm}%
\protect\raisebox{-12.5pt}[0pt][0pt]{\protect\includegraphics[height=.8cm]{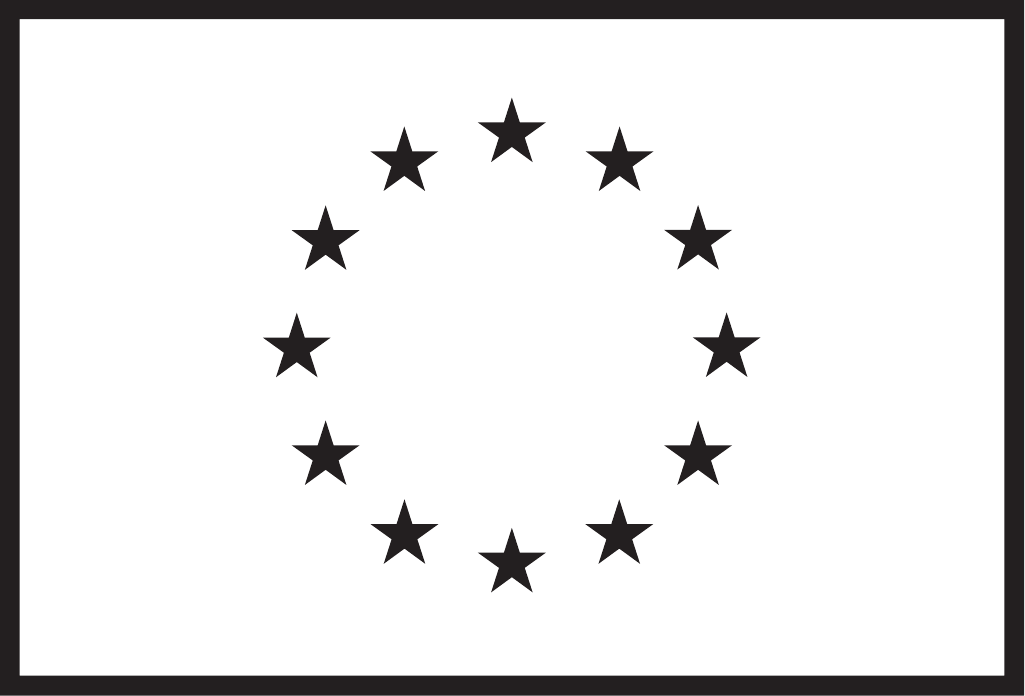}}%
\protect\end{wrapfigure}%
This project has received funding from the European Union's Horizon 2020 research and innovation programme under grant agreement no. 786698. The work reflects only the authors' view and the Agency is not responsible for any use that may be made of the information it contains.}}

\author{\IEEEauthorblockN{Irina Baptista\IEEEauthorrefmark{1},\vspace*{4pt} Stavros Shiaeles\IEEEauthorrefmark{1} and Nicholas Kolokotronis\IEEEauthorrefmark{2}}

\IEEEauthorblockA{\IEEEauthorrefmark{1}Centre for Security, Communications and Networks Research (CSCAN), Plymouth University, Plymouth PL4 8AA, UK\\
Email: {\tt irina.baptista@plymouth.ac.uk}, {\tt stavros.shiaeles@plymouth.ac.uk}}

\IEEEauthorblockA{\IEEEauthorrefmark{2}Department of Informatics and Telecommunications, University of Peloponnese, 22131 Tripolis, Greece\\
Email: {\tt nkolok@uop.gr}}
}

\maketitle

%----------------------------------------------------------%
%                      Section Change                      %
%----------------------------------------------------------%
\begin{abstract}
The continued evolution and diversity of malware constitutes a major threat in modern systems. It is well proven that security defenses currently available are ineffective to mitigate the skills and imagination of cyber-criminals necessitating the development of novel solutions. Deep learning algorithms and {\em artificial intelligence} (AI) are rapidly evolving with remarkable results in many application areas. Following the advances of AI and recognizing the need for efficient malware detection methods, this paper presents a new approach for malware detection based on binary visualization and self-organizing incremental neural networks. The proposed method's performance in detecting malicious payloads in various file types was investigated and the experimental results showed that a detection accuracy of $91.7$\% and $94.1$\% was achieved for ransomware in .pdf and .doc files respectively. With respect to other formats of malicious code and other file types, including binaries, the proposed method behaved well with an incremental detection rate that allows efficiently detecting unknown malware at real-time.
\end{abstract}

%----------------------------------------------------------%
%                      Section Change                      %
%----------------------------------------------------------%
\begin{IEEEkeywords}
Security; malicious software; machine learning; self-organizing neural networks; binary visualisation.
\end{IEEEkeywords}

%----------------------------------------------------------%
%                      Section Change                      %
%----------------------------------------------------------%
\section{Introduction}
\label{sec:intro}

\IEEEPARstart{T}{he} increasing sophistication and diversity of malware has long ago surpassed the capabilities of malware analysts. Vast amounts of data need to be analyzed every day in search of potential threats, but the current techniques are simply not enough to comply with the high demand \cite{kpgram18}. Malware writers use different methods, such as polymorphism and obfuscation, to outperform detection systems. Typical approaches often depend on signature-based detection, where a file's signature is compared to a list of known malicious ones. Though such approaches are computationally less demanding, they are susceptible to simple evasive techniques and zero-day exploits. Common detection techniques rely on static and dynamic analysis. Static analysis is computationally efficient and safer than other methods as it does not require the code's execution to allow being analyzed; however, it is also easily affected by evasive techniques. On the other hand, dynamic analysis has higher chances of detecting unseen samples but heavily affects the system performance since the file is analyzed in a virtual environment.

To overcome the above-mentioned problems, in this paper we investigate and evaluate an efficient detection mechanism relying on visual representations of malware. The proposed approach utilises binary visualisation, converting a file's binary data into an image, and {\em self-organizing incremental neural networks} (SOINN) for the analysis and detection of malicious payloads. By doing so, the likelihood of detecting obfuscated code in the file is increased \cite{malin12}. The algorithm chosen for the binary representation uses Hilbert space-filling curves to cluster the data, ensuring that the data is grouped in an optimal way. Moreover, the use of SOINN in the detection mechanism, which is an unsupervised machine learning algorithm known for its incremental abilities, provides the ability of learning fast by exploiting only the information needed for building the {\em neural network} (NN) and removing redundant nodes. These properties of SOINN are inherited from {\em self-organizing maps} (SOM) and {\em growing neural gas} (GNG) \cite{shen06}. Our contribution is the novel combination of binary visualization with SOINN whose robustness to noise makes it lightweight and faster than other approaches, and hence ideal for very large datasets allowing the provisioning of efficient detection mechanisms as a service. A prototype was developed for analyzing detection performance, where malware samples from various categories and file types were included to test the model in different scenarios. The experimental results showed a high overall detection accuracy, which attained its maximum value $94.1$\% for particular malware types.

The paper is organized as follows: in Section \ref{sec:related} we discuss related work in the area of binary visualization and machine learning, as well as other approaches combining both methods. Sections \ref{sec:method}, \ref{sec:results} describe the proposed methodology and experimental results. Finally, a discussion of the results obtained is presented in Section \ref{sec:discussion} before concluding in the last section.

%----------------------------------------------------------%
%                      Section Change                      %
%----------------------------------------------------------%
\section{Related work}
\label{sec:related}

Analysis and detection of malware have been a persistent challenge for security professionals. Traditional attempts are focused on static and dynamic analysis, but the rapid growth and evolution of malicious code have compelled researchers in order to derive novel analysis and detection solutions. {\em Machine learning} (ML) is among the innovative technologies that have been employed towards that direction. Many works have focused on building frameworks for analysis \cite{gavrilut09,kolter04,nissim14}, acquiring static features \cite{islam10,uppal14} and classifying malware families \cite{lakhotia13,pirscoveanu15,rieck11}. A system relying on ML to detect an unknown malicious payload, without the need for removing obfuscation, was presented in \cite{kolter04}; text classification methods were shown to improve the detection accuracy of obfuscated samples. A comparison of various ML algorithms, e.g. naive Bayes, random forest and {\em support vector machine} (SVM), on detecting malicious {\em application programming interface} (API) call sequences was perfomed in \cite{uppal14}, where the superiority of the SVM classifier was demonstrated. The main weakness of ML-based detection systems is that they rely on a virtual environment to analyze samples; this not only affects their run-time performance, but also endangers the whole system since the samples need to be executed. In addition, the ability of malware to adapt its behavior to the execution environment, the excessive number of features that need to be extracted per sample, and the high volume of malware instances being reported, reduce the chances of accurate detection \cite{shiaeles12,shiaeles14}.

Alternative approaches exploring the use of samples' visual representation have been proposed; most efforts consider the clustering of malware by families or similarity \cite{gove14,han14a,long14,saxe12}, whereas other works focus on evaluating the sample-to-image transformation with the majority utilizing grayscale images \cite{han15,han14b}. Examples of solutions proposed include the {\em visualization of executables for reversing \& analysis} (VERA) framework \cite{quist09}, which represents on a 3-dimensional space a program's flow to identify obfuscating parts of its code, and the {\em similarity evidence explorer for malware} (SEEM) tool \cite{gove14}, which provides the ability to visually compare a given binary with classified samples to identify the similar ones. Although this line of research is quite promising, certain challenges need to be tackled. More precisely, most approaches are based on malware classification and not on its detection that is still performed by existing tools. In addition, the proposed approaches are often not robust to commonly used techniques that attackers could use, like data relocation and data redundancy, to circumvent the detection mechanism \cite{han14b}.

Techniques investigating the combination of both aforementioned categories for malware analysis have not received much attention. Computer vision techniques and SVMs were used in \cite{kancherla13} for malware detection achieving an accuracy of $95$\% on a dataset with 37K samples. A combination of ML, visualization and steganalysis was proposed in \cite{burgess14} to detect packed malware; both SVMs and a variation of the {\em $k$-nearest neighbour} (KNN) were used, with the later achieving a classification accuracy of $99.5$\%. A similarity detection framework with 1.2M samples was used in \cite{kirat13} for malware classification; an accuracy of $99$\% was achieved (for about half of the samples), but the use of similarity considerably lowers the probability of detecting unknown malware with dissimilar structure. Unfortunately, the above approaches are also susceptible to common attackers' techniques \cite{nataraj11}. Thus, it seems that binary representation and ML still pose challenges, and resemble more than a modern addition to current systems rather than a replacement.

%----------------------------------------------------------%
%                      Section Change                      %
%----------------------------------------------------------%
\section{The proposed method}
\label{sec:method}

\begin{figure}[t]
\centering
\includegraphics[width=\linewidth]{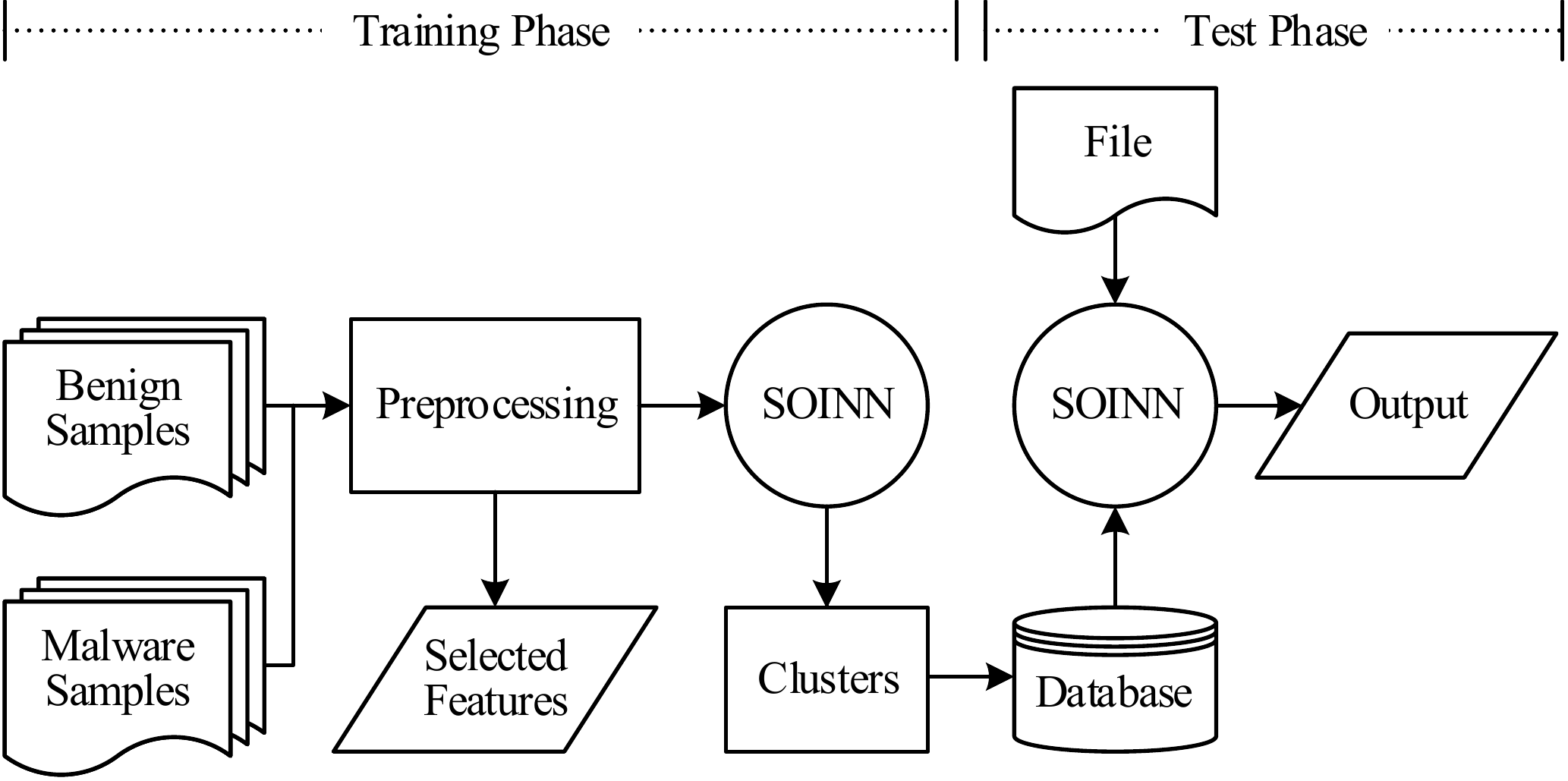}
\caption{The system architecture is separated in two stages. The first stage is the training phase, where the samples are processed and the topological structure of the NN is built. In the second stage the files are tested against the samples in the database to perform classification.}
\label{fig.architecture}
\end{figure}

This section presents the methodology used for developing the proposed malware detection system illustrated in Fig. \ref{fig.architecture}; it relies on visualizing a file's binary onto a two-dimensional space and then using SOINN, after having performed feature extraction, to distinguish between benign and malicious files. A dataset of binary images was first generated from the sample files (Section \ref{sec:visualisation}). The images are next pre-processed, where prominent features containing valuable information about the image are automatically extracted to form a vector of length $1024$ (Section \ref{sec:feature}). The vectors are then provided to SOINN (Section \ref{sec:soinns}), which performs clustering and classification; the training data are stored in a database for subsequent use during analysis and detection. A similar process is followed during the testing phase where, after having pre-processed the input file and generated the feature vector, the outcome is evaluated against the training data. The selection of the nearest vector (to the input given) and the class type is based on the Euclidean distance.

%----------------------------------------------------------%
%                      Section Change                      %
%----------------------------------------------------------%
\subsection{Binary visualisation}
\label{sec:visualisation}

\begin{figure}[t]
\centering
\begin{subfigure}{0.49\linewidth}
\includegraphics[width=\textwidth]{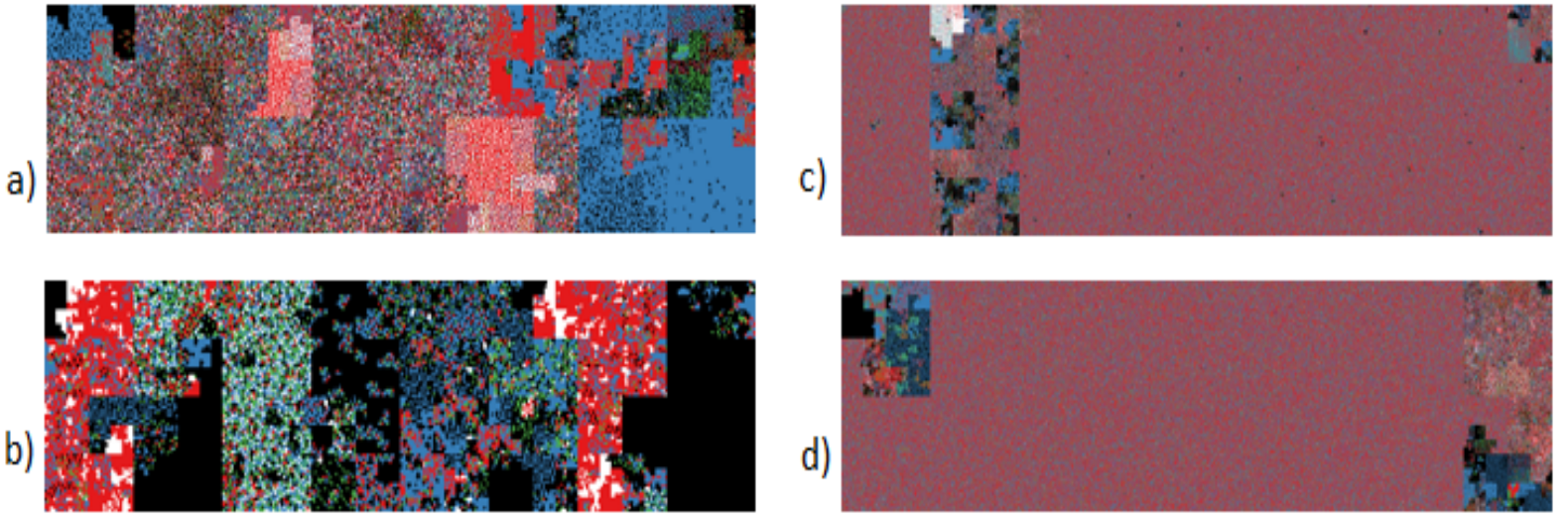}
\caption{{\tt Backdoor.Win32.Shoda bot.b}\vspace{6pt}}
\end{subfigure}
\hfill
\begin{subfigure}{0.49\linewidth}
\includegraphics[width=\textwidth]{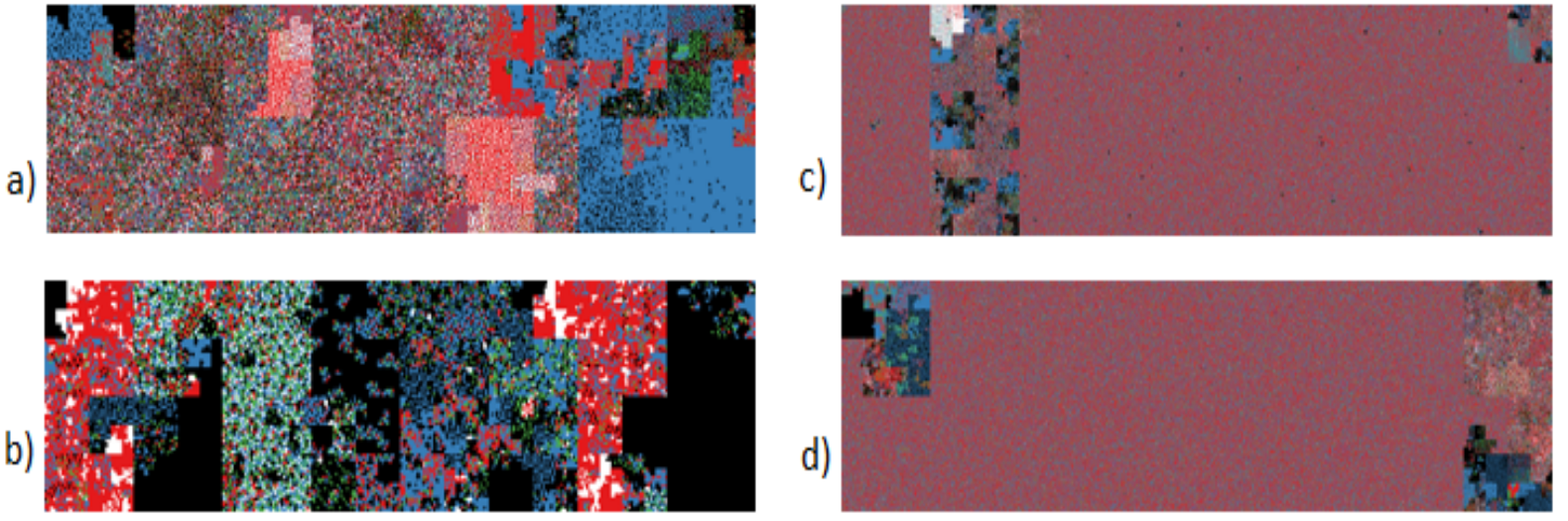}
\caption{{\tt Trojan-Dropper.Win32 .HeliosBinder.p}\vspace{6pt}}
\end{subfigure}
\newline
\begin{subfigure}{0.49\linewidth}
\includegraphics[width=\textwidth]{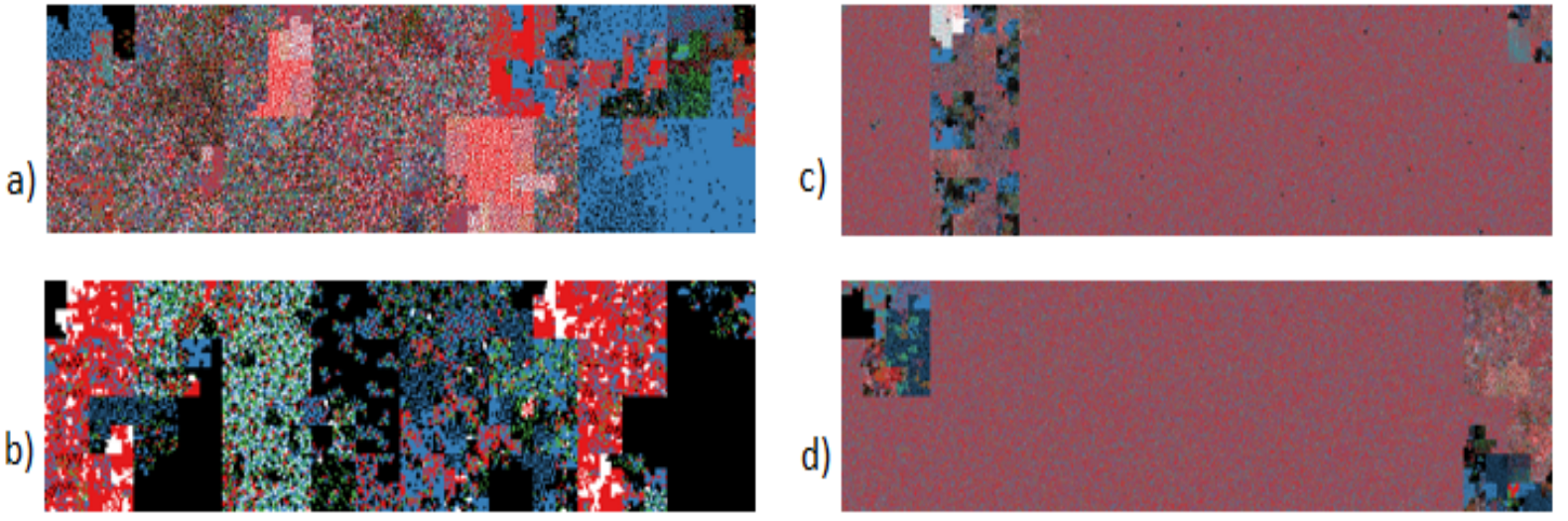}
\caption{{\tt Vmware player}}
\end{subfigure}
\hfill
\begin{subfigure}{0.49\linewidth}
\includegraphics[width=\textwidth]{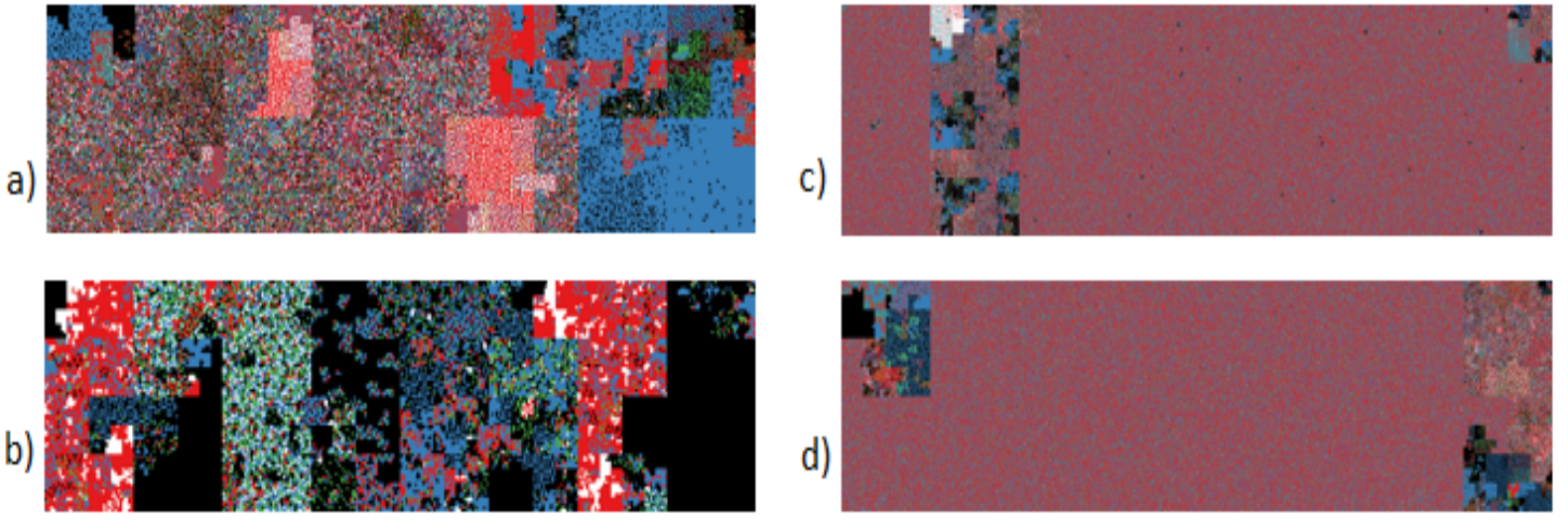}
\caption{{\tt Google}}
\end{subfigure}
\caption{Visual representations of malicious (a), (b) and benign (c), (d) executable samples.}
\label{fig.examples}
\end{figure}

Binary visualization transforms the binary contents of a file to another domain that can be visually represented (normally a 2-dimensional space). Our visual representation algorithm is based on {\tt binvis.io} \cite{cortesi16}, an online tool that uses colour schemes to represent different binary or ASCII values. The clustering algorithm that is employed in Fig. \ref{fig.architecture} is based on the Hilbert space-filling curve, which outperforms other curves in preserving the locality between objects in multi-dimensional spaces \cite{faloutsos89,jagadish97}, thus creating a much more appropriate imprint of the image. The above combination is useful for detecting obfuscated code in malicious files, which could go unnoticed otherwise, since they tend to be grouped together. To convert a file into an image, its data are seen as a byte string, where each byte's value is compared against the ASCII table and is attributed to a color according to the division it belongs:
\begin{itemize}
\item {\em blue} color,  if the character is printable;
\item {\em green} color, if the character is control; and
\item {\em red} color,   if the character is extended.
\end{itemize}
Besides the above divisions in the ASCII table, two more classes were added, namely \hex{00} (colored {\em black}) and \hex{FF} (colored {\em white}), to represent null and (non-breaking) spaces respectively ---an example visualization is given in Fig. \ref{fig.examples}.

%----------------------------------------------------------%
%                      Section Change                      %
%----------------------------------------------------------%
\subsection{Feature extraction}
\label{sec:feature}

During the pre-processing stage important features, capable of indicating the presence of malicious payload, are extracted from the image so as to be used in the classification process. The images' size is an important factor for the algorithm's performance. Indeed, by having images as small as $28\times 28$ pixels, a vector of length $784$ is obtained that might just not be large enough to include meaningful information for the distinction of malicious and benign files. On the other hand,  this might still be the case even when the size is increased to $128\times 128$ pixels. It is thus reasonable to assume that not all the values are relevant to distinguish the classes, since specific locations and patterns carry more information than other image areas. To have a valid analysis and increase the probability of detecting peculiarities, it is necessary to first detect the {\em regions of interest} (RoI) on the image where specific malware patterns may appear.

Similar to our methodology, \cite{sun14} is clustering images using SOINN, but the work concerned the identification of persons over non-overlapping cameras. In order to obtain a reduced number of features, while minimizing the impact on system's performance, images were divided into $6$ stripes to account for regions having different characteristics, since this was found to provide improved results during identification \cite{park06}; different colour spaces and texture features were utilized on each stripe prior to conversion into histograms \cite{gray08,zheng13}, thus significantly reducing the feature vector's length.

The same approach is adopted in our case for obtaining the feature vector from the images, with the difference of having the images  divided into $4$ parts ---top, bottom, upper and lower middle--- and using only the RGB color space to compose the histogram. These regions were chosen due to the differences occurring in the majority of patterns obtained from each. The overall histogram, which serves as the feature vector of length $k = 1024$, is the result of concatenating the individual regions' histograms, each having length $256$ (the RGB color space). It was shown during the experimentation that the number of features did not have a noticeable impact on the classification accuracy or the system performance.

%----------------------------------------------------------%
%                      Section Change                      %
%----------------------------------------------------------%
\subsection{Self-organizing incremental neural network}
\label{sec:soinns}

The feature vectors extracted from the images are given to SOINN, which has two layers; the first layer aims at learning the topological structure of the NN, whereas the second layer determines the number of clusters based on the input received from the first layer. Among other things, SOINN maintains a set $\mathcal{N}$ of nodes, the set $\mathcal{C}$ of connections between the nodes, and the set $\mathcal{W}$ of weights (i.e.\ feature vectors) associated with each node. The algorithm used for training the NN is employed at both layers, with the difference that the similarity threshold $T$ determining if a new node should be added in $\mathcal{N}$ is adaptive (resp.\ constant) at the first (resp. second) layer. The algorithm is briefly described below; the details can be found in \cite{shen06}.

The algorithm is initialized by setting $\mathcal{N} = \{n_1, n_2\}$, with weights $\mathcal{W} = \{\vec{w}_1, \vec{w}_2\}$ chosen uniformly at random from the feature vectors, and $\mathcal{C} = \emptyset$ to be the empty set. For each input vector $\vec{u} \in \mathbb{R}^k$, the algorithm computes the closest nodes
\begin{align}
\ell^{\color{white}\prime} &= \argmin_{i \in \mathcal{N}} \eucnorm{\vec{u} - \vec{w}_i} \\
\ell' &= \argmin_{i \in \mathcal{N} \setminus\! \{\ell\}} \eucnorm{\vec{u} - \vec{w}_i}
\end{align}
referred to as the {\em winner} and {\em second winner} respectively, based on the Euclidean distance metric $\delta(\vec{x}, \vec{y}) = \eucnorm{\vec{x} - \vec{y}}$. If no similarity exists between the input vector $\vec{u}$ and either $\vec{w}_{\ell}$ or $\vec{w}_{\ell'}$, the new node is inserted into the network $\mathcal{N}$; otherwise, a connection $(\ell,\ell')$ between the winners is added to $\mathcal{C}$, if not present, and its {\em age} is set to zero, i.e.\ $\alpha(\ell,\ell') = 0$. Next, the age of all connections with $\ell$ as a starting point is increased by $1$. Since this process would create regions of highly connected nodes (corresponding to the formed classes), the algorithm is using an age threshold $A$ in order to determine if a connection $(i,j) \in \mathcal{C}$ must be dropped. More precisely, if we have that a connection $(i,j)$ is such that $\alpha(i,j) > A$, then its importance is low compared to others and can be omitted. The similarity threshold used (at the first layer) above is node-specific, and is determined by the following function
\begin{equation}
T_i = \max_{j \in \mathcal{L}_i} \eucnorm{\vec{w}_i - \vec{w}_j} \,, \quad \forall\, i \in \mathcal{N}
\end{equation}
where $\mathcal{L}_i$ is either the set of neighbors of node $i$, if this set is nonempty, or it is equal to $\mathcal{N} \setminus \{i\}$ otherwise.

Noise removal is one of the important aspects of SOINN that separates it from other algorithms and essentially allows the NN to maintain only the valuable pieces of information. To eliminate noisy nodes the algorithm is first provided with a $\lambda$ value that determines the number of iterations the algorithm has to go through before searching for nodes that do not give important data. If a node has no neighbors, this means there is a possibility of distorting the pattern; nodes with two or less neighbors are removed only if considered insignificant. After the NN has been completely trained, the Euclidean distance is applied again to classify a file; the argument corresponding to the shortest distance is recognized as the winning node and its class is used to classify the input vector.

%----------------------------------------------------------%
%                      Section Change                      %
%----------------------------------------------------------%
\section{Experimental results}
\label{sec:results}

In this section, we present the performance analysis results of the prototype that was implemented based on the above methodology. The dataset that was used contained a total of $2$K %$1362$
benign files, obtained from trusted, common and portable applications, and a total of $2$K %$2834$
malicious files collected from the {\tt VirusShare} website and included classes like viruses, worms, backdoors, rootkits, and trojans. Both the benign and malicious samples admitted the same distribution into various file types (with a predominance of executable files), which is illustrated in Table \ref{tab.distribution}.

\begin{table}[h]
\centering
\caption{Malware distribution according to file types; ``other'' refers to DoS, packed, e-mail and exploit.}
\label{tab.distribution}
\setlength{\tabcolsep}{5pt}
\renewcommand{\arraystretch}{1.1}
\begin{tabular}{lcccccc}
\hline
         & {\tt.exe}                & {\tt.doc}                & {\tt.pdf}& {\tt.txt}& {\tt.htm}& Total \\
\hline
Virus    & ${\color{white}0}1.37$\% & $14.57$\%                &          & $3.28$\% & $1.03$\% & ${\color{white}1}20.25$\% \\
Worm     & ${\color{white}0}1.22$\% &                          &          & $0.43$\% &          & ${\color{white}10}1.65$\% \\
Backdoor & $25.02$\%                &                          &          & $0.63$\% &          & ${\color{white}1}25.65$\% \\
Trojan   & $39.44$\%                & ${\color{white}0}0.42$\% &          & $0.96$\% & $2.39$\% & ${\color{white}1}43.21$\% \\
Rootkit  & ${\color{white}0}2.65$\% &                          &          &          &          & ${\color{white}10}2.65$\% \\
Other    & ${\color{white}0}2.63$\% &                          & $2.43$\% & $0.66$\% & $0.87$\% & ${\color{white}10}6.59$\% \\
\hline
Total    & $72.33$\%                & $14.99$\%                & $2.43$\% & $5.96$\% & $4.29$\% & $100.00$\% \\
\hline
\end{tabular}
\end{table}

\begin{figure}[t]
\centering
\includegraphics[scale=.51]{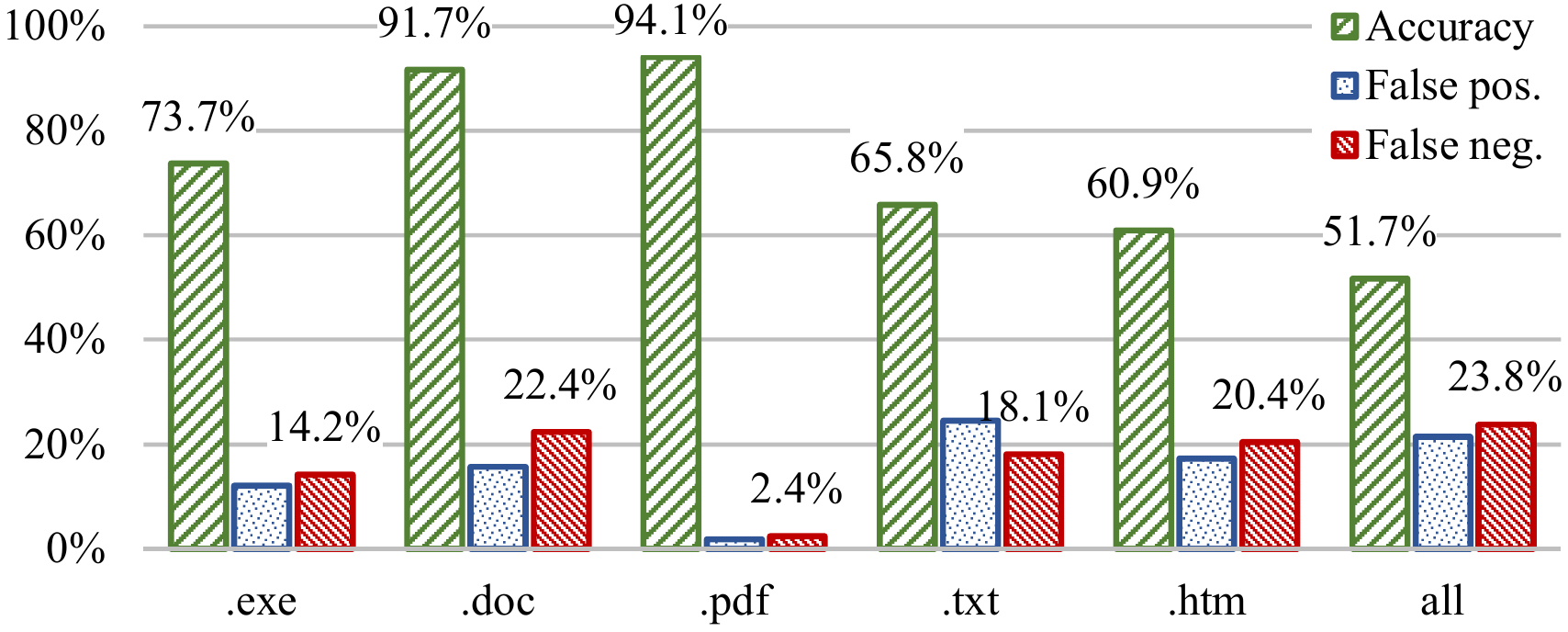}
\caption{Accuracy rate and the number of false positives/negatives observed during testing for various file types.}
\label{fig.accuracy}
\end{figure}

\begin{figure}[t]
\centering
\includegraphics[scale=.51]{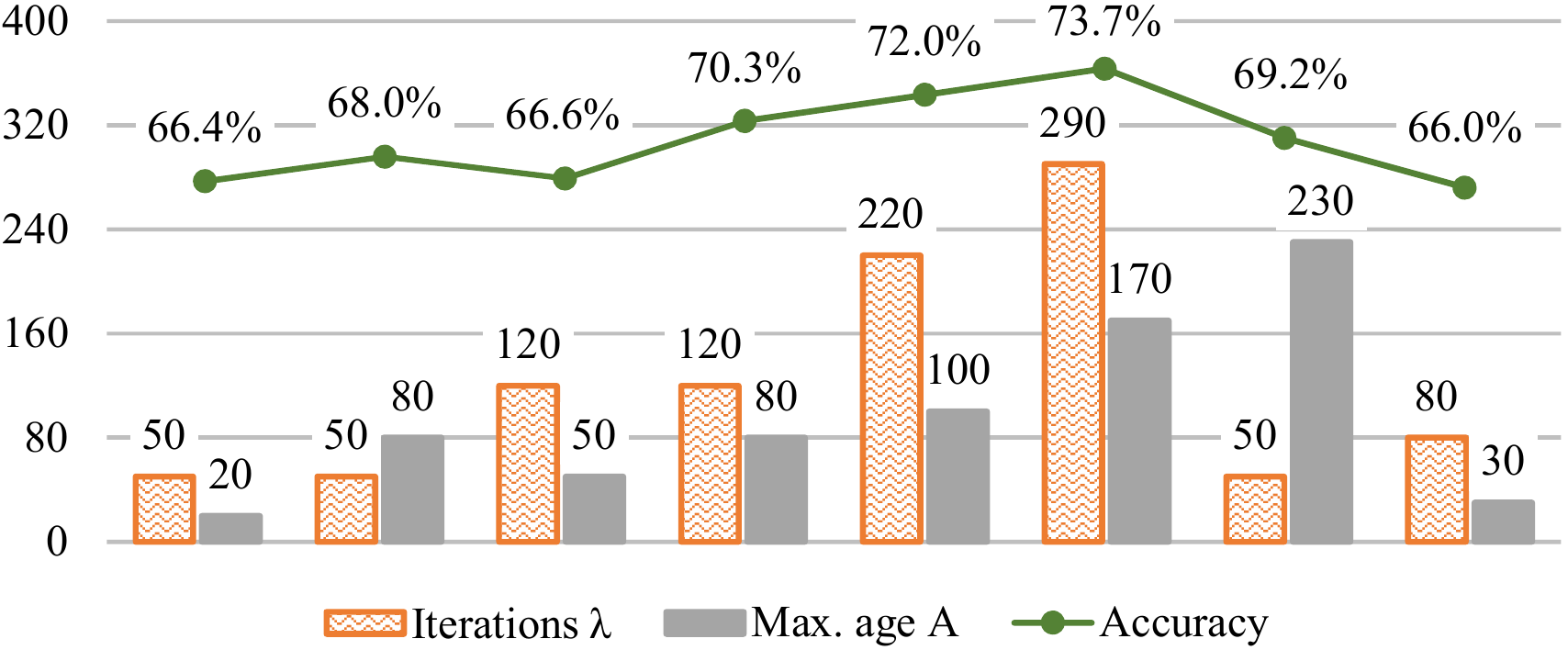}
\caption{The influence of parameters $\lambda,A$ on the accuracy rate. The line shows the variation of the accuracy rate according to the parameters' values.}
\label{fig.influence}
\end{figure}

The performance of SOINN was assessed for various values of the parameters $\lambda$ and $A$ in terms of accuracy rate. For each setup, $100$ Monte Carlo simulations were performed, with the same dataset, in order to get the average accuracy rate and the number of {\em false positives} (FP) and {\em false negatives} (FN); the results obtained are shown in Fig. \ref{fig.accuracy}. The algorithm behaves well in distinguishing malicious from benign files in all cases with the exception of text files ({\tt.txt} and {\tt.htm}). The fact that benign {\tt.doc} and {\tt.pdf} files have less number of clusters in the images than the malicious ones (this was marked by the algorithm as a feature for malware), along with the fact that they do not just contain printable ASCII characters, has improved detection performance. A general remark is that the strong differences in the format of files do not allow obtaining a single solution fitting all cases, therefore having an impact on system's performance. A solution to this problem is to take a file's type into account, allowing the algorithm to properly adjust its behavior in each case. The impact that $\lambda$ and $A$ have on accuracy performance is shown in Fig. \ref{fig.influence}. It is seen that increasing $A$ excessively has an adverse effect in removing noise from the NN, and the time that the system needs for processing information increases with $\lambda,A$. The best accuracy rate during testing was achieved for $\lambda = 290$ and $A = 170$. The training of the system was extremely fast, taking about $15$ seconds to compute a dataset with $4$K feature vectors.

\begin{figure}[t]
\centering
\includegraphics[scale=.51]{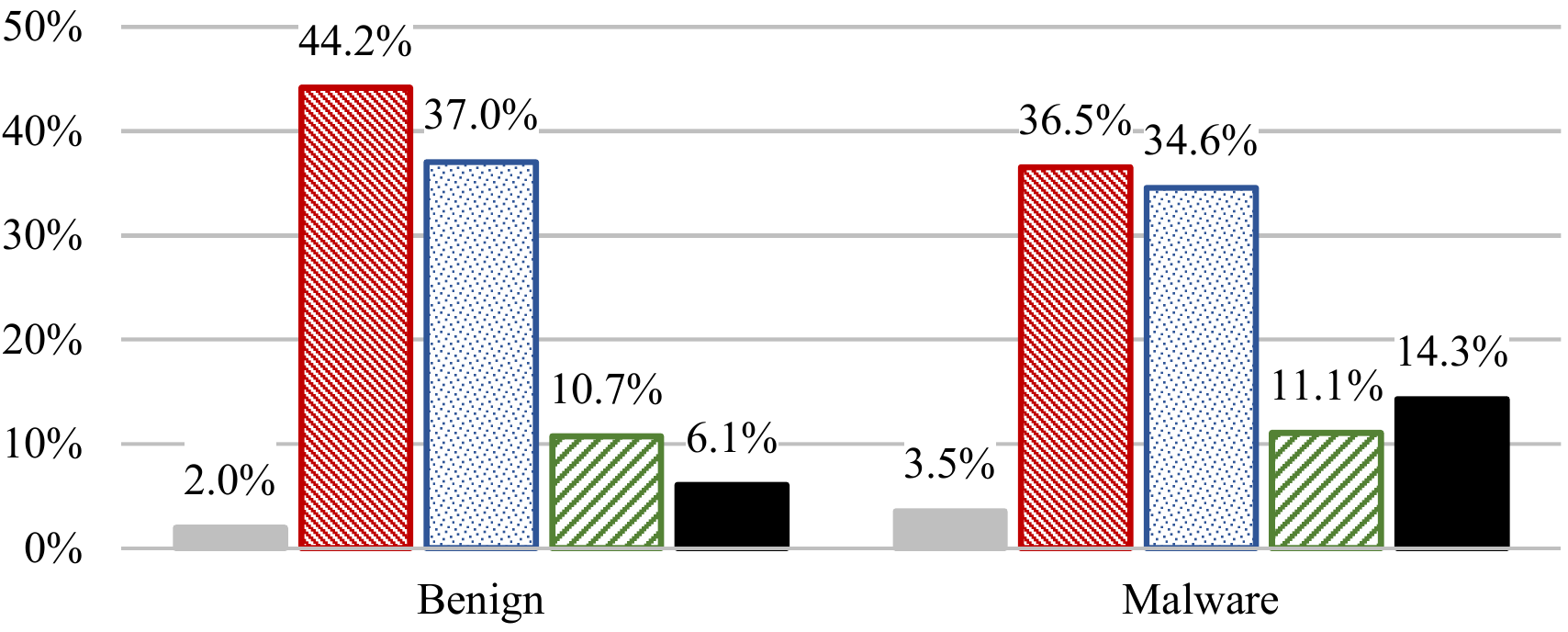}
\caption{Average color frequency and influence on the classification between malware and software}
\label{fig.colour}
\end{figure}

Moving to binary visualization, the images acquired reveal that it is possible to identify malware files through images. By representing a file's binary contents as an image, the potential patterns and discrepancies in data start to emerge. In Fig. \ref{fig.examples}, the binary information composing examples of malicious and benign files is shown. Malicious files have a tendency for often including ASCII characters of various categories, presenting a colorful image, while benign files have a cleaner picture and distribution of values. Another technique to identify malicious files is to search for green and black areas since malware has a high predominance of control and null characters. In contrast, benign files are mostly made of printable and ASCII extended characters. A further aspect to note is that most benign files have a few clusters at the top and bottom of the image whereas malicious files have a higher variety of clusters. An insight on the color distribution of each class (i.e.\ benign {\em vs.}\ malware) is given in Fig. \ref{fig.colour}. Although the green color is observed in about the same percentage of pixels, their spreading in the images differs with malware admitting regions of high concentration. The difference in black pixels is evident, since the use of null characters to transport and hide data or to exploit a system is quite common \cite{dadkhah14,wressnegger16}.

An experiment was first performed, based on the dataset of $4$K samples, to evaluate SOINN's detection accuracy and to obtain a quick estimate of its performance. To have more accurate information about its behavior when presented with high data volumes, another dataset of $200$K samples was also created. The algorithm was found to be highly efficient and the time needed to complete the training phase of the last dataset was under $15$ minutes in a personal workstation equipped with an Intel Core i5 processor.

%----------------------------------------------------------%
%                      Section Change                      %
%----------------------------------------------------------%
\section{Discussion}
\label{sec:discussion}

Using visual representation techniques allows to obtain an insight into the structural differences between malicious and benign files. Such differences are harder to identify in some file types; to be more precise, although {\tt .txt} files are mostly composed of printable characters (blue), this is not the case in a (large) number of samples due to the existence of control characters, something that also holds for infected {\tt .txt} files thus resulting in increased difficulty during analysis. To solve this problem, more representative (of a malware's features) samples could be used having a mixture of colors and clusters in the image. Executable files exhibit a more diverse color distribution as they include different categories from the ASCII table. Hence, in contrast to text files, it is unlikely that a high percentage of blue pixels will appear in a benign executable file's image representation, something that in turn increases the chances of being malware. Due to the above, it is evident why training the same NN for all file types drastically decreases the detection accuracy rate by $22$\%. This however could be overcome if the files' extension is also taken as input during the analysis and training of the neural network. By analyzing Fig. \ref{fig.accuracy}, it is readily seen that the proposed algorithm is better in identifying text-based files than other file types. This means that our approach can accurately detect ransomware that has recently become one of the preferred and most consequential forms of attack. To the best of our knowledge, no other work has tested the proposed methodology for different file types; instead, solely executables are being used for measuring the performance of a detection mechanism.

Based on our experiments and the colour distribution, it was observed that malicious samples had a larger number of black colored  regions than benign files. These regions represent null characters and various motives exist behind their use when the malware is written. Such characters, apart from representing the end of a string or being used for padding \cite{huseby05,sikorski12}, are also used by cyber-criminals to hide data or instructions in the registry \cite{ligh11,sovarel05}, a practice sought after by malware writers; however, it equally likely that these characters indicate (possibly malicious) unknown files that are being created at a host \cite{dahl13}. Several attacks are possible just by using null characters, e.g.\ to circumvent security mechanisms and load an infected file in order to allow remotely controling a system \cite{fonseca10}. More details about the use of null characters for launching cyber-attacks can be found at the website of {\em common attack pattern enumeration and classification} (CAPEC) community. Regarding the patterns created, malware images tend to have more compact clusters and a variety of colors, whereas benign files are generally represented by almost static images with only a few and small-scale clusters. It was also noticed that the top and bottom half of the images carry information that is more crucial to the analysis and detection of malware than the rest of the areas, since this is where different clusters, from picture to picture, are often perceived.

\begin{figure}[t]
\centering
\includegraphics[scale=.51]{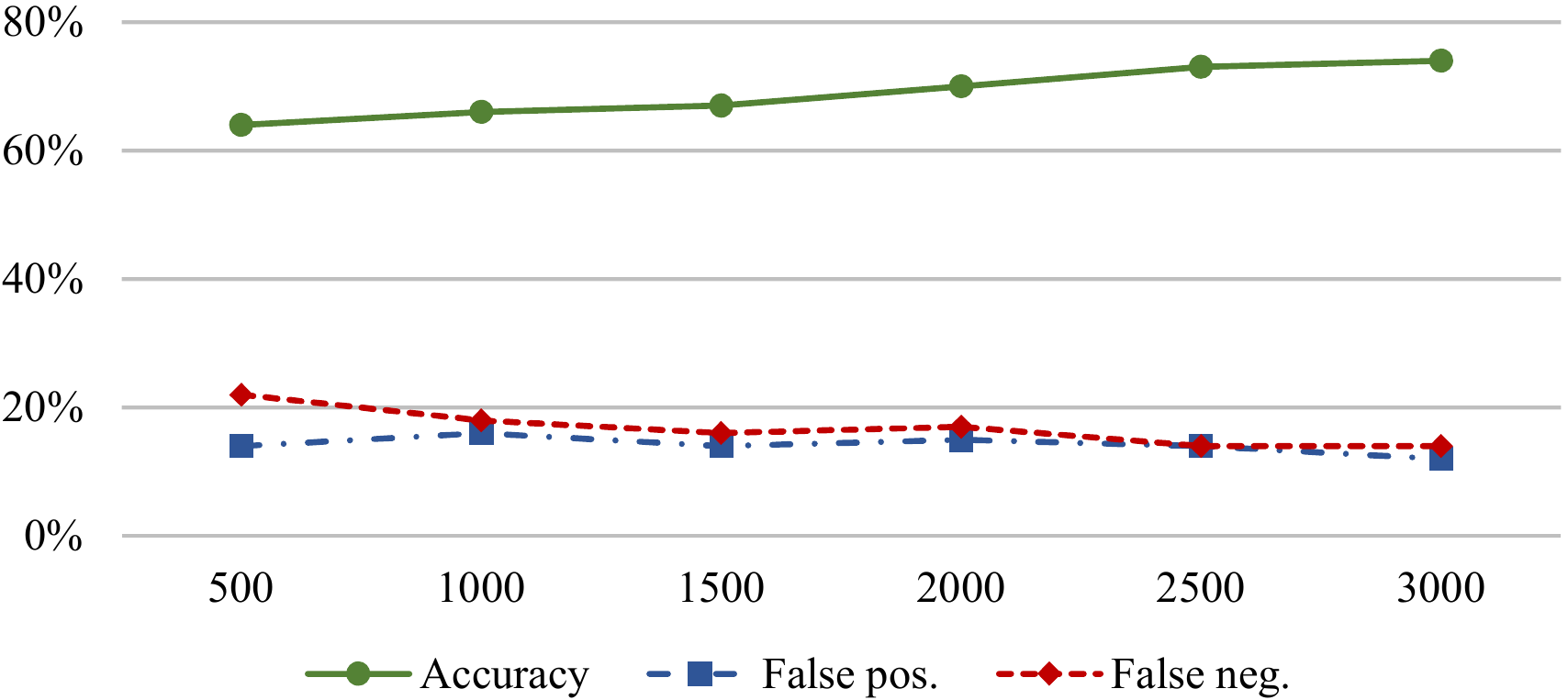}
\caption{Accuracy increase as a result of the increase in the amount of samples.}
\label{fig.samples}
\end{figure}

Regarding the algorithm's performance, SOINN achieved an accuracy of $73.7$\% for just over $4$K samples. As Fig. \ref{fig.samples} shows, the accuracy rate (resp. false positives and false negatives) increases (resp. decrease) with the number of samples, and thus it would be possible to get much better results provided that more samples are added during the training phase. SOINN is capable of reducing its internal structure without having to abdicate from new or old knowledge (stability-plasticity dilemma). Its noise removal properties ensure that the network is not limited in capacity, a problem that is found in {\em nearest neighbours} algorithm \cite{kirat13}. SVMs have similar drawbacks; the major problem are the highly complex and extensive memory requirements, as well as, the slow response during the testing. Ont the other hand, SOINN is lightweight and extremely fast, the whole network was trained in under $15$ seconds for about  $10$K iterations.

\begin{figure}[t]
\centering
\includegraphics[scale=.51]{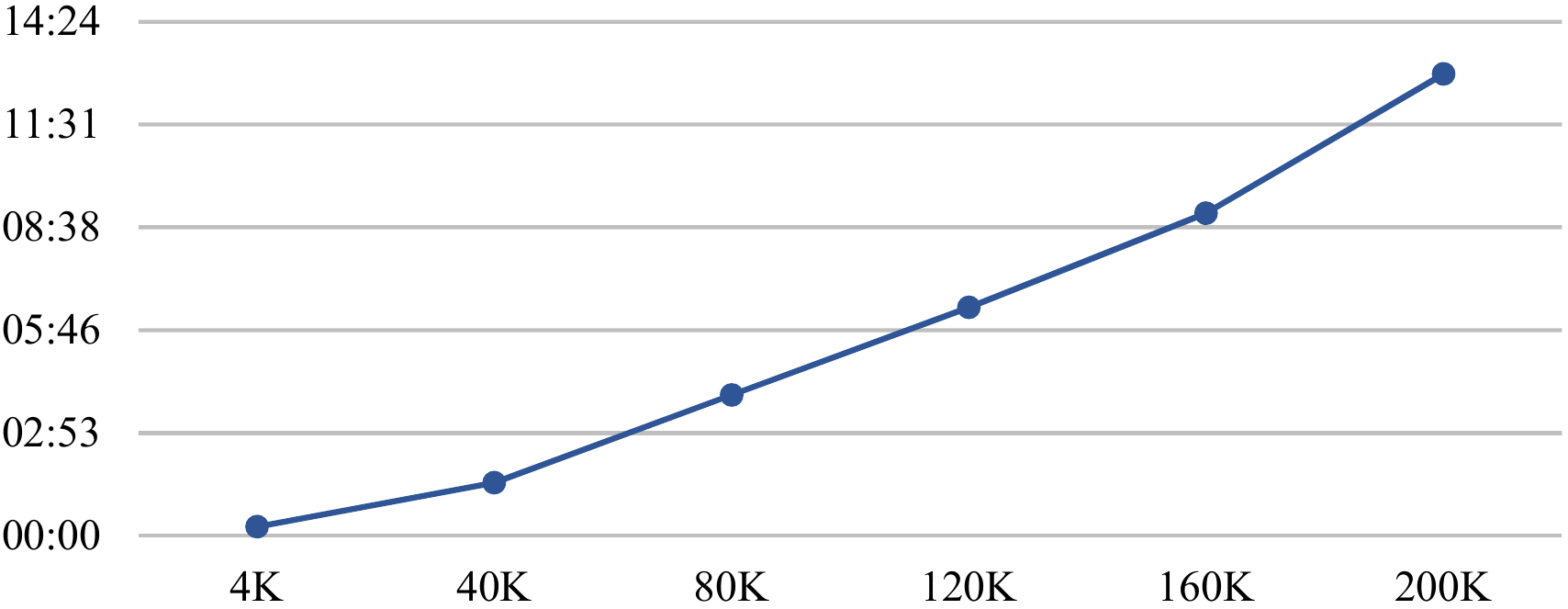}
\caption{Training phase of the algorithm with different volumes of data (x-axis) and the time (minutes:seconds) needed to process the data (y-axis).}
\label{fig.training}
\end{figure}

%----------------------------------------------------------%
%                      Section Change                      %
%----------------------------------------------------------%
\section{Conclusions}
\label{sec:conclusion}

In this paper, an image-based malware detection technique was proposed that uses unsupervised learning. The study tested if malicious files could be differentiated from benign files by focusing on features extracted from their visual representation. The proposed approach utilizes feature vectors of length $1024$ and does not employ any computationally intensive technique or algorithm, which allowed fast processing and training times. The experimental tests were conducted for different file types and malware categories and revealed that the algorithm, whose accuracy is improved with the available number of samples, can be successfully used for malware detection. In particular, the algorithm achieved an overall average detection rate of about $74$\% with $12$\% false positives and $14$\% false negatives. Ongoing work, aims at extending the approach presented to mobile platforms as a lightweight cloud-based antivirus. Since there is great potential in improving the results obtained by enhancing feature extraction, a number of options is currently being considered, including Gabor texture filters and multi-resolution analysis techniques among others.

%----------------------------------------------------------%
%                      Section Change                      %
%----------------------------------------------------------%

\end{document}